\documentclass[twocolumn]{aastex701}
\usepackage{amsmath}

\shorttitle{Joint Line Intensity Recovery from Multi-Tracer Cross-Correlations}
\shortauthors{Sierra \& Karkare}

\begin{document}

\title{Joint Line Intensity Recovery from Multi-Tracer Cross-Correlations}

\author[orcid=0000-0002-9246-5571] {Carlos E. Sierra}
\affiliation{Department of Physics, Stanford University, 382 Via Pueblo Mall, Stanford, CA 94305, USA}
\affiliation{SLAC National Accelerator Laboratory, 2575 Sand Hill Road, Menlo Park, CA 94025, USA}
\affiliation{Kavli Institute for Particle Astrophysics and Cosmology, Stanford University, 452 Lomita Mall, Stanford, CA 94305, USA}
\email[show]{csierra@stanford.edu}

\author[orcid=0000-0002-5215-6993] {Kirit S. Karkare}
\affiliation{Department of Physics, Boston University, 590 Commonwealth Avenue, Boston, MA 02215, USA}
\email{kkarkare@bu.edu}

\begin{abstract}
Line intensity mapping (LIM) provides a unique window into the high-redshift universe, but the separation of signals from overlapping emission lines and bright continuum foregrounds remains a central analysis challenge. At millimeter wavelengths in particular, [CII], [CI] and numerous CO transitions overlap with comparable brightness, leading to significant spectral confusion. This work presents a joint likelihood framework that incorporates multi-tracer cross-correlations both between distinct LIM tracers and with external galaxy surveys. By jointly fitting all auto- and cross-spectra between the LIM and galaxy data, this approach combines the spatial and spectral information in each dataset to break degeneracies between overlapping lines and recover individual bias-weighted line intensities at fixed cosmology, even at redshifts where no direct cross-correlation is available. We validate this framework on realistic mocks of an extended SPT-SLIM-like survey spanning 125--175~GHz over 100~deg$^2$, combined with an LSST Y1-like photometric galaxy catalog, and recover CO(2-1) through CO(6-5) and [CI](1-0) over $0.3\lesssim z\lesssim4$ in the presence of bright continuum foregrounds. We demonstrate that adding the photometric galaxy data tightens line constraints by a factor of $\sim$3--4 where the galaxies overlap the lines in redshift, and by $\sim$20--50\% at higher redshifts beyond the galaxy coverage, where the gain comes indirectly from constraining the low-redshift interlopers. These results highlight external tracer overlap as a primary consideration when optimizing the frequency coverage and footprint of near-future LIM experiments.

\end{abstract}

\section{Introduction}
The detection of faint emission lines from high-redshift galaxies through line intensity mapping (LIM) offers a complementary window to traditional galaxy surveys for studying cosmic structure formation \citep{Lidz2011,Bernal2022}. Unlike conventional approaches that trace structure by resolving individual source galaxies, LIM measures the aggregate emission from all galaxies within an observing pixel, enabling lower-resolution but highly-efficient observations of the distant high-redshift universe where single-object detections become prohibitively expensive. The maps produced by these observations probe the history of cosmic star formation~\citep{Bolatto2013,Carilli2013,Tacconi2020}, trace the evolution and morphology of reionization~\citep{Gong2011,Pritchard2012,Yue2015}, and place cosmological constraints on the expansion history, primordial non-Gaussianity, and large scale structure at redshifts inaccessible to galaxy surveys~\citep{Karkare2018,Bernal2019a,MoradinezhadDizgah2019,Bernal2019b,MoradinezhadDizgah2022}. However, realizing these science goals hinges on cleanly separating the target signal from contaminant emission, where the dominant obstacles are line confusion from interlopers and bright continuum foregrounds.

Several strategies have been proposed to mitigate interloper contamination. Pixel masking can remove spatial regions dominated by bright low-redshift interlopers, either blindly based on pixel brightness~\citep{Visbal2011, Breysse2015} or guided by external galaxy catalogs~\citep{Sun2018}. However, aggressive masking has been shown to discard substantial information content~\citep{Karoumpis2024} and introduces its own biases if the mask geometry is not properly accounted for in the power spectrum estimation. 

Cross-correlation with external tracers offers a more robust alternative by exploiting the fact that different redshift slices have distinct clustering signatures. This approach has been studied extensively, both in the context of cross-correlations with discrete galaxy surveys~\citep{Switzer_2019,Chung2019, VanCuyck2026} and cross-correlations between different spectral lines originating from the same redshift~\citep{Visbal2010,Roy_2024,Cheng_2024,McBride2024}. In both cases, the cross-spectrum between two tracers is sensitive only to the redshift where their kernels overlap, suppressing the contribution of interlopers that originate from different redshifts. 

The effectiveness of this approach depends on the availability of the external (or internal) datasets and the statistical framework used to combine them. Cross-correlations with galaxy surveys are most effective at low redshifts $z\lesssim 2$, where the distribution of galaxy samples tends to be most dense. On the other hand, line-line cross-correlations, which can probe significantly higher redshifts than galaxy surveys allow, are most powerful when multiple transitions from the same redshift are simultaneously accessible, either within a single wide-band survey or across complementary LIM experiments observing at different wavelengths.

In this work, we demonstrate a multi-tracer approach that exploits the clustering statistics of both line intensity maps and galaxy surveys to break the degeneracies inherent in interloper-contaminated spectra. By constructing a joint likelihood that incorporates line-line ($I_{a}I_{b}$), galaxy-line ($gI$), and galaxy-galaxy ($gg$) power spectra, we show that cross-correlation measurements provide the additional constraints needed to recover individual line amplitudes in the presence of many interlopers. This approach builds on previous work in the area of LIM cross-correlations \citep[e.g.][]{Switzer_2019, Feng_2019, Roy_2024, Cheng_2024} but extends the framework to jointly solve all correlated emission in multi-tracer datasets. The strategy parallels the joint multi-probe analyses now standard in cosmic microwave background (CMB) and large-scale structure cosmology, where combining auto- and cross-power spectra of multiple tracers into a single likelihood breaks parameter degeneracies that would not be resolvable from any individual probe \citep[e.g.][]{Krause2017,Farren2024,Adame2025}.

This joint likelihood framework can serve as a critical tool for optimizing the survey design of future LIM experiments. By quantifying the information content inherent in different cross-correlation combinations, we can inform the strategic selection of survey parameters such as frequency coverage, sky area, and depth. For instance, this framework allows us to identify optimal frequency bands that maximize the number of LL cross-correlations (e.g. observing CO(2-1) and CO(4-3) at the same redshift) while minimizing the impact of the brightest foregrounds. One may also assess the trade-offs between small, deep-field observations versus a wide-field shallow survey, using the signal-to-noise of the recovered line amplitudes as a quantitative metric. This is particularly relevant in the context of cross-correlations with external surveys, where the sensitivity to the cross-power signal is often mode-limited rather than noise-limited. Because the variance of the cross-correlation signal scales inversely with sky area, this approach is ideal for optimizing spatial overlap with external tracer catalogs in a LIM survey.

This paper is organized as follows: in Section 2, we give an overview of the multi-tracer power spectrum formalism. Section 3 details the mock datasets built from realistic simulations, including observational systematic effects and foreground removal methods. In Section 4, we present the joint likelihood framework that incorporates clustering statistics from line intensity mapping and galaxy datasets to fit multiple spectral lines simultaneously. Section 5 describes the results of applying this framework to mock SPT-SLIM and LSST datasets, and Section 6 summarizes our conclusions.
We demonstrate that the joint likelihood approach is robust at separating many overlapping emission lines of comparable strength, recovering individual line amplitudes that would be inaccessible from any individual dataset alone.

\section{Multi-Tracer Power Spectrum}
In a complete halo model framework, the total power spectrum of a line intensity field is expressed as the sum of three primary components
\begin{equation}
    P_\mathrm{tot}(k,z) = P^{2h}(k,z) + P^{1h}(k,z) + P_\mathrm{shot}(z) \ ,
\end{equation}
where $P^{2h}$ is the 2-halo term representing clustering between distinct halos, $P^{1h}$ is the 1-halo term describing the distribution of emitters within individual halos, and $P_\mathrm{shot}$ is the Poisson shot noise arising from the discrete nature of the sources. On the angular scales considered in this work ($\ell < 1000$), the shot noise term is scale-independent, but the clustering is not purely linear. In the mock data, we find that non-linear clustering contributes non-negligibly above $\ell\sim400$ (Figure~\ref{fig:templates}). At low redshift $z\approx0.5$, this excess is well described by the non-linear matter power spectrum from the halofit model~\citep{Takahashi2012}, while at higher redshifts it exceeds the halofit prediction, indicating an additional contribution from the 1-halo term or scale-dependent bias of the line emitters. We therefore fit the shot noise and the scale-independent part of the 1-halo term together as a single noise floor $N_i(z)$, and absorb any remaining non-linear clustering into the empirical templates described in Section~\ref{sec:templates}.
\begin{equation}
    N_i(z) \equiv P^{1h}_i(z) + P_\mathrm{shot}^i(z) \ ,
\end{equation}
treating $N_i$ as a free parameter fitted jointly with the clustering amplitude. For two intensity fields $I_i$ and $I_j$, where $i,j$ index the set of target emission lines, the 2-halo term is modeled as
\begin{equation}
    P^{2h}_{ij}(k,z) = \langle bI\rangle_i (z) \, \langle bI\rangle_j (z) \, P_m(k,z) \ ,
\end{equation}
where $\langle bI\rangle_i(z) \equiv b_i(z)\bar{I}_i(z)$ is the bias-weighted mean intensity of line $i$, the primary parameter we constrain in this analysis, and $P_m(k,z)$ is the linear matter power spectrum. Under the Limber approximation, the angular power spectrum $C_\ell$ for a narrow redshift slice at comoving distance $\chi(z)$ is related to the 3D power spectrum by:
\begin{equation}
    C_\ell^{I_i I_j} = \frac{1}{\chi^2(z)\,\Delta\chi(z)} \,
    \langle bI\rangle_i(z)\,\langle bI \rangle_j(z)\,
    P_m\!\left(k_\ell(z),\, z\right) \ .
\end{equation}
Here $k_\ell(z) \equiv (\ell+1/2)/\chi(z)$ and $\Delta\chi(z)$ is the comoving
depth of the redshift slice, so that the $1/\chi^2\Delta\chi$ prefactor carries
any 3D power spectrum into its angular counterpart.
In practice, a single LIM frequency channel $\nu$ receives contributions from multiple emission lines at different redshifts, with each line $i$ mapping to $\nu_\mathrm{obs}$ via $1+z_i = \nu_{\mathrm{rest},i}/\nu_\mathrm{obs}$. The total observed auto-power spectrum is therefore a sum over all contributing lines:
\begin{equation}
\begin{split}
    C_\ell^{\nu\nu} = \sum_{i} \frac{1}{\chi^2(z_i)\,\Delta\chi(z_i)} \bigg[
    & \langle bI \rangle_i^2(z_i)\, P_m\!\left(k_\ell(z_i),\, z_i\right) \\
    & {}+ N_i(z_i) \,\bigg] \ .
\end{split}
\label{eq:auto}
\end{equation}
This sum highlights the fundamental degeneracy in the LIM auto spectrum between the line signal of interest and foreground interlopers. Proposed strategies for dealing with this interloper problem include pixel masking \citep{Karoumpis2024}, redshift-space distortions \citep{Lidz2016}, template fitting \citep{Cheng2020}, and cross-correlations. In this work, we focus on the cross-correlation approach and consider the general case of incorporating correlations between multiple tracer types. 

First, cross-correlations between different LIM frequency channels $\nu$ and $\nu'$ isolate power from pairs of lines $i$ and $j$ that originate from the same redshift $z$, such that $z_i(\nu) = z_j(\nu') = z$. The line-line cross-spectrum is modeled as

\begin{equation}
\begin{split}
    C_\ell^{\nu\nu'} = \frac{1}{\chi^2(z)\,\Delta\chi(z)} \bigg[
    & \langle bI \rangle_i(z)\,\langle bI \rangle_j(z)\, P_m\!\left(k_\ell(z),\, z\right) \\
    & {}+ \rho_{ij}\,\sqrt{N_i(z)\,N_j(z)} \,\bigg] \ .
\end{split}
\label{eq:llcross}
\end{equation}
We note that the clustering term assumes perfect correlation between the two line tracers, i.e. a clustering cross-correlation coefficient
\begin{equation}
    r_{ij} \;\equiv\; \frac{P_{ij}^{2h}(k,z)}{\sqrt{P_{ii}^{2h}(k,z)\,P_{jj}^{2h}(k,z)}} \;=\; 1 \ ,
    \label{eq:rij}
\end{equation}
where $i$ and $j$ index either a line or a galaxy sample. Strictly speaking, this assumption is not generally true. As explained by \cite{Schaan2021}, decorrelation is expected on small scales where the power spectrum is dominated by the 1-halo and shot noise components. In these regimes, the signal is sensitive to so-called ``line noise,'' which characterizes the fractional covariance of line luminosities driven by stochastic processes within individual galaxies and halos. This scatter in luminosity thus prevents perfect correlation between two line tracers. However, we find in the mock data that the correlation between two lines at the same redshift remains close to unity out to $\ell=1000$, with the small decline at high $\ell$ driven by the shot noise term rather than any decorrelation at clustering scales (Appendix~\ref{sec:decorrelation}). This is expected, since $N_i$ includes the 1-halo contribution and is therefore sensitive to the line noise described above. We thus fit for a shot correlation coefficient $\rho_{ij}$ for every pair of lines at the same redshift, with a flat prior over $[0,1]$.

Second, cross-correlations with an external tracer of known redshift, e.g., a galaxy survey, are sensitive to the overlap in redshift space between the galaxy selection and each line's emission. Denoting the linear galaxy bias and redshift overlap kernel as $b_g(z_g)$ and $W(z_i, z_g)$ respectively, the galaxy-line cross-spectrum ($g\nu$) is given by
\begin{equation}
\begin{split}
    C_\ell^{g\nu} = {} & \frac{1}{\chi^2(z_g)\,\Delta\chi_g} \,
    \langle bI \rangle_i(z_i)\, b_g(z_g) \\
    & \times P_m\!\left(k_\ell(z_g),\, z_g\right)\, W(z_i, z_g) \ .
\end{split}
\label{eq:glcross}
\end{equation}

Once again, we assume perfect correlation between the galaxy and line clustering, which we find to be well justified in the mock data over the overlapping redshift range $0.3<z<1.6$ (Appendix~\ref{sec:decorrelation}). We do not include a separate shot noise term in the $g\nu$ cross-spectrum, since any galaxy-line shot noise in the mocks is already contained in the empirical templates (Section~\ref{sec:templates}). Finally, combining the galaxy auto-spectrum,
\begin{equation}
    C_\ell^{gg} = \frac{1}{\chi^2(z_g)\,\Delta\chi_g}
    \left[ b_g^2(z_g)\,P_m\!\left(k_\ell(z_g),\, z_g\right)
    + N_g(z_g) \right] \ ,
    \label{eq:ggauto}
\end{equation}
with the $g\nu$ cross-spectrum breaks the $\langle bI \rangle$--$b_g$ degeneracy in Eq.~\ref{eq:glcross}, where $N_g=1/\bar{n}$ is the Poisson shot noise of the galaxy sample. 

\section{Mock Datasets}
\label{sec:mocks}
\begin{figure}
    \centering
    \includegraphics[width=0.9\linewidth]{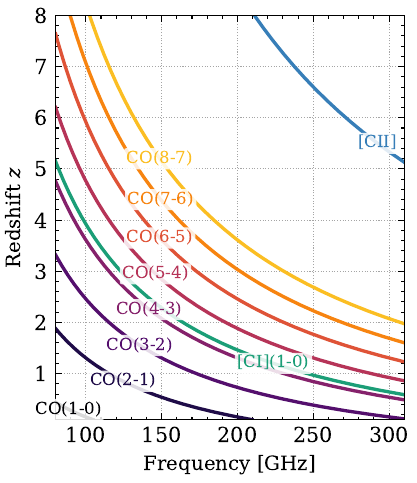}
    \caption{Rest-frame frequencies of [CII], [CI], and the CO rotational ladder mapped to observed frequency as a function of redshift. Multiple transitions fall within the same mm-wave band at different redshifts, highlighting the line confusion problem.}
    \label{fig:ladder}
\end{figure}
\subsection{LIM Cube Mocks}
We focus our application of the multi-tracer framework on the millimeter-wave band, where line confusion from overlapping [CII], [CI], and CO transition lines is high (Figure~\ref{fig:ladder}). We consider a millimeter-wavelength LIM experiment that observes these lines over a wide redshift range, with the goal of isolating the contribution of each line to the observed power spectrum. The line rest frequencies are known, and an optical galaxy survey covering the same sky area provides redshift information for a subset of the underlying sources. Recovering the individual line amplitudes requires a mock dataset that supplies both the millimeter-wave intensity field and the correlated optical galaxy distribution generated from a common underlying matter field so that the cross-correlation signal between the two tracers is preserved. Although many experimental configurations could be tested within this framework, we focus on a configuration that approximates the SPT-SLIM~\citep{Karkare2022} and LSST Y1~\citep{LSST_DESC} survey specifications.

To generate the mock line intensity mapping data cubes, we use the SIDES~\citep{Bethermin2017,Bethermin2022} simulation code. Starting from an input halo light cone, SIDES produces millimeter-wave spectral intensity cubes incorporating continuum CIB and spectral lines as tracers of large-scale structure. The CIB is empirically modeled using redshift-evolving spectral energy distributions (SEDs) calibrated against far-IR/sub-millimeter data from \textit{Herschel} to reproduce observed continuum number counts and power spectra. For spectral lines, SIDES derives luminosities for the CO rotational ladder, [CII] and [CI] based on infrared luminosities and star formation rates. The model incorporates environmental effects by linking CO excitation to the mean UV radiation field and accounts for gravitational lensing magnification on line fluxes. Additionally, SIDES distinguishes between main-sequence and rarer starburst galaxy populations, assigning separate empirical scaling relations for both SED selection and CO-to-IR correlations to each galaxy type. To ensure realistic stochasticity, the model incorporates an intrinsic scatter of approximately 0.2 dex across all line-luminosity scaling relations. We note that the mean line brightness temperatures in the resulting cubes are consistent with the recent empirical estimates of the mean CO and [CII] brightness from \citet{Chiang2026}.

The light cone used in this work is adapted from \citet{SatoPolito2023} and uses halo catalogs from the MultiDark Planck 2 N-body simulations~\citep{MDPL2}. Dark matter halos are populated with star formation rates and stellar masses via the \texttt{UniverseMachine}~\citep{Behroozi_2018} prescription. From this light cone product, we generate 16 independent $10^\circ \times 10^\circ$ sky realizations with redshifts extending out to $z\sim10$. The ensemble mean over these realizations defines the model templates of Section~\ref{sec:templates}, and the fits presented below are run on a single realization. We find that the mm-wave clustering properties are consistent between this catalog and the 117~deg$^2$ SIDES-Uchuu light cone catalog described in \citet{Gkogkou2023}.

The mock LIM observations span $125<\nu<175$~GHz across 51 channels of 1~GHz width each ($R\sim150$), covering a $100 \deg^2$ sky area at $\sim1$ arcmin angular resolution. This configuration approximates the design of the SPT-SLIM pathfinder~\citep{Karkare2022}, an on-chip spectrometer concept deployed on the 10~m South Pole Telescope during the 2024-2025 austral summer season. Comparable mm-wave LIM experiments include TIME \citep{Butler2025}, CONCERTO \citep{Fasano2022}, and CCAT \citep{CCATPrimeCollaboration2022}, which probe overlapping frequency ranges with different combinations of bandwidth, spectral resolution, and sky coverage. Instrument noise is modeled for each channel assuming background-limited dual-polarization spectrometers observing a $\sim15$~K atmosphere-dominated sky at 150~GHz with an optical efficiency of 0.2. This results in an effective NET $\sim$~530 $\mathrm{\mu K \sqrt{s}}$ for a single 150 GHz channel.

We also add a lensed CMB field to the mock cubes. At these frequencies, the CMB is a second bright continuum that varies smoothly with frequency, and including it is an additional robustness check that the cleaning in Section~\ref{sec:cib} is not tuned to the CIB alone. We generate a single lensed sky with \texttt{pysm3} and interpolate it onto each tile's pixel grid, scaling a single 150~GHz slice to the other channels by converting from thermodynamic to Rayleigh-Jeans brightness temperature.

\begin{deluxetable}{ll}
\tablecaption{Baseline mock survey configuration.\label{tab:config}}
\tablehead{
\colhead{Parameter} & \colhead{Value}
}
\startdata
\multicolumn{2}{c}{\textit{LIM survey (SPT-SLIM-like)}} \\
\hline
Frequency range            & 125--175~GHz                  \\
Channel bandwidth          & $d\nu = 1$~GHz                \\
Number of channels         & 51                            \\
Sky area                   & 100~deg$^2$ (87~deg$^2$ effective)\tablenotemark{a} \\
$\mathrm{NET_{det}}$\tablenotemark{b}      & 710-780~$\mu$K\,$\sqrt{\rm s}$ across the band \\
Number of feeds    & 200              \\
Observing time             & 1~yr                         \\
\hline
\multicolumn{2}{c}{\textit{Photometric galaxy survey (LSST Y1-like)}} \\
\hline
Galaxy density            & $\bar n\sim18$~gal/arcmin$^2$ \\
Redshift range             & $0.2 < z < 1.6$                \\
Redshift uncertainty          & $\sigma_z = 0.03\,(1+z)$      \\
Tomographic bin width      & $3\sigma_z(1+z_0)$             \\
Number of $z$ bins         & 9                            \\
\hline
\enddata
\tablenotetext{a}{Effective area after apodization, which sets $f_\mathrm{sky}$ in the covariance.}
\tablenotetext{b}{$\mathrm{NET_{det}}$ refers to the noise-equivalent temperature of a single detector. Here we assume on-chip filter-bank spectrometers that consist of a dual-polarization feed coupled to two detectors per spectral channel (one per polarization). The array sensitivity per spectral channel is then given by $\mathrm{NET_{arr}}=\mathrm{NET_{det}}/\sqrt{2 N_\mathrm{feeds}}$. }
\end{deluxetable}

\begin{figure*}
    \centering
    \includegraphics[width=\linewidth]{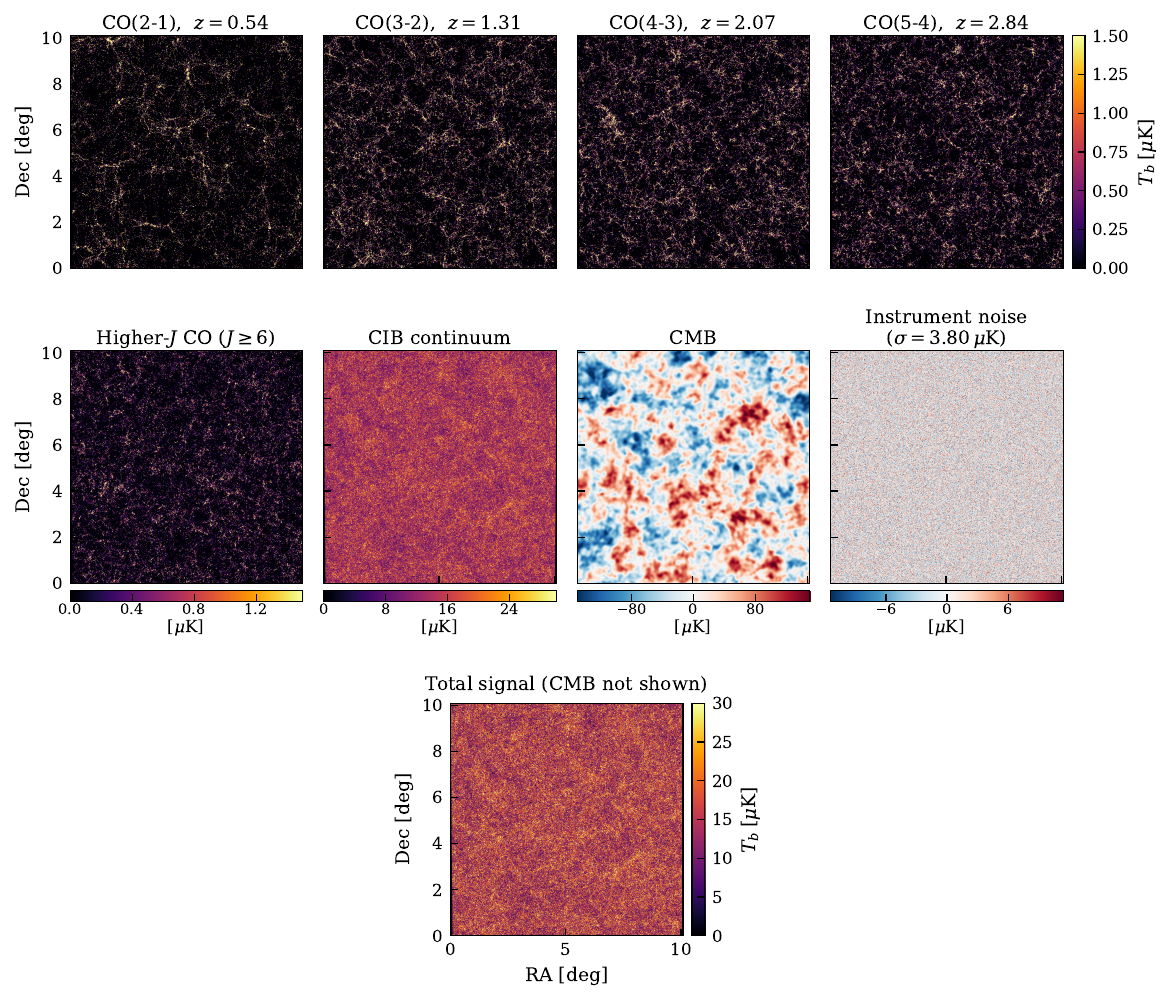}
    \caption{Mock line intensity mapping observations simulated at 150~GHz for a 100 $\deg^2$ field using the SIDES~\citep{Bethermin2022} code. The mocks include 9 CO transitions, [CI], and [CII], as well as the CIB and CMB continuum foregrounds and instrument noise that assumes $N=200$ background-limited SPT-SLIM-style spectrometers observing for one year. The goal of this work is to recover the individual line components in the top row from the ``Total signal'' dataset (CMB not shown) containing interlopers, bright foregrounds, and realistic instrument noise.}
    \label{fig:sims}
\end{figure*}

\subsection{Galaxy Catalog Mocks}
The mock galaxy survey datasets are synthesized from the same underlying light cone catalog used for the line intensity mapping mocks through an effective abundance matching selection process. We use the redshift distribution $dN/dz$ from the Dark Energy Survey Year 3 Gold data set as observed over the SPT-Deep Field as a baseline template, and scale the density to match the projected depth of the Rubin Observatory Legacy Survey of Space and Time (LSST) Year 1 gold sample \citep{LSST_DESC}. To simulate photometric redshift (photo-$z$) error, a Gaussian smear is applied to the true galaxy redshifts:
$$z_{\text{phot}} = \mathcal{N}(z_{\text{true}}, \ \sigma_z(1+z_{\text{true}})) \ , $$
assuming a baseline scatter of $\sigma_z = 0.03$ following the LSST Y1 requirements. The galaxy sample is tomographically binned into widths $3\sigma_z(1+z_0)$ such that the bulk of the smeared signal is captured within each slice, therefore mitigating the impact of the photo-$z$ uncertainty. 

To match the target survey density, we keep only the most massive galaxies in each redshift bin, setting a stellar mass threshold $M_\mathrm{cut}$ and discarding galaxies with $M_{\star} < M_\mathrm{cut}$ to yield the target $dN/dz$. The selected galaxies are projected onto a 2D spatial grid with pixel resolution matching the line intensity mapping survey. This yields redshift-binned overdensity maps
$$\delta_g = \frac{N - \langle N \rangle}{\langle N \rangle} \ ,$$ 
where $N$ is the galaxy count in a given pixel and $\langle N \rangle$ is the sky-averaged galaxy count for that redshift bin. 

\subsection{Power Spectrum Estimation}
\label{sec:FT}
Angular power spectra are computed using the small-sky Fourier transform approximation implemented via the \texttt{pixell} library. For the limited sky areas considered in this work, we find this approach to be computationally efficient and sufficiently accurate, showing no significant bias compared to more sophisticated pseudo-$C_\ell$ methods such as MASTER.

To compute the cross-spectrum $C_\ell^{ab}$ between two maps $m_a$ and $m_b$, we first apply a spatial apodizing window function $W(\mathbf{x})$ to mitigate the effects of spurious mode coupling and data artifacts at the edges of the survey geometry. The sky-averaged monopole is subtracted from each map prior to applying the window taper to suppress the $W(\mathbf{x}) \times \text{monopole}$ term from generating spurious power at large scales. The Fourier-space representation of the windowed maps is given by $\tilde{m}(\mathbf{\ell}) = \text{FT}\{m(\mathbf{x})W(\mathbf{x})\}$, and the raw 2D cross-power spectrum is estimated as
$$P_{ab}(\mathbf{\ell}) = \frac{1}{w_2} \text{Re}\{\tilde{m}_a(\mathbf{\ell})\tilde{m}_b^*(\mathbf{\ell})\} \ ,$$ 
where $w_2 = \langle W^2(\mathbf{x}) \rangle$ accounts for the power loss after applying the window taper. We bin the 2D power spectrum into five logarithmically-spaced 1D angular bins over the range $100<\ell<1000$. 

\subsection{Continuum Foreground Removal}
\label{sec:cib}
The dominant correlated foreground in mm-LIM is the CIB, which behaves as a smooth signal in frequency and is significantly brighter than any individual spectral line on the sky across all spatial scales (Figure \ref{fig:sims}). Moreover, both the CIB and emission lines trace the same underlying matter field and are therefore spatially correlated, making the foreground difficult to mitigate via masking or cross-correlation techniques. Additionally at these frequencies, the CMB itself adds a second smooth contaminant background that is about an order of magnitude brighter than the CIB fluctuations (Figure~\ref{fig:sims}). However, this is offset by the fact that it is uncorrelated with the line signals and has a frequency dependence that is known exactly and identical in every pixel.

One option for dealing with the continuum is to model it as a separate component in the likelihood. This is the standard approach for CMB power spectrum analyses, where the CIB spectral energy distribution is treated as a thermal modified blackbody that spatially traces the linear power spectrum \citep{Reichardt2021,Louis2025}. In our joint likelihood, however, this approach would require additional modeling of the correlation between the CIB and every line and galaxy bin in the data vector.

In this work, we remove both the CIB and CMB emissions from the LIM data cube, $T(\nu)$, by least-squares fitting and subtracting a second-order polynomial along the line-of-sight direction of each voxel. In other words,

\begin{equation}
    T_\mathrm{clean}(\nu) = T(\nu) - \sum_{n=0}^2 \alpha_n \nu^n \ .
\end{equation}

\begin{figure}
    \centering
    \includegraphics[width=\linewidth]{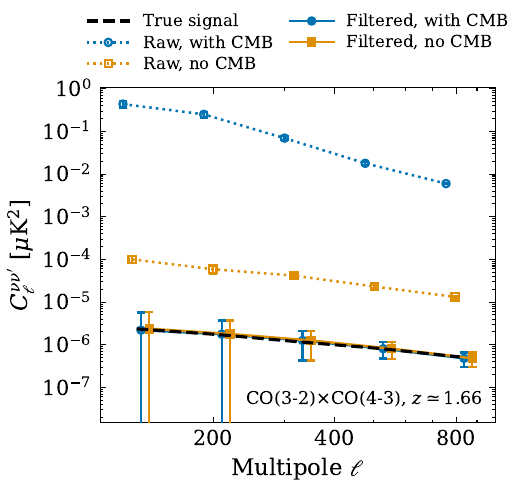}
    \caption{Effects of continuum removal on the mock dataset for an example line-line cross-spectrum. The correlated CIB foreground, which is roughly 50x brighter than the target signal in the cross-spectrum, is removed from the data alongside the much brighter CMB background by fitting a second-order polynomial along the line-of-sight, as described in Section \ref{sec:cib}. The residual line signal loss is effectively recovered by computing the expected transfer function calibrated against reasonable mock cubes. We demonstrate this removal on mocks with and without the CMB component (blue and orange curves, respectively). The filtered spectra and error bars remain unchanged at the 1\% level, so the CMB leaves no measurable imprint on the cleaned data.}
    \label{fig:cib-clean}
\end{figure}

This simple approach has been shown to be an effective way of removing smoothly varying foregrounds from intensity mapping datasets, particularly for analyses of 21~cm neutral hydrogen~\citep{Liu2011,Alonso_2015,Gkogkou_2026}. Figure \ref{fig:cib-clean} demonstrates the effectiveness of this approach in completely removing the continuum foreground power from the contaminated dataset. As expected, the filtering can also remove some of the power from the line signal of interest. We correct for this signal loss by multiplying the angular template of every filtered pair by an analytic transfer function $F_\ell^{ab}$. Since the fit is linear in the data and the fitted continuum is subtracted at every channel, the cleaning is a fixed linear operation along the frequency axis. Arranging each voxel's spectrum into a vector, the cleaned spectrum is $\mathbf{T}_\mathrm{clean} = \mathbf{M}\mathbf{T}$, where the operator
\begin{equation}
    \mathbf{M}=\mathbf{I}-\mathbf{X}(\mathbf{X}^\top\mathbf{X})^{-1}\mathbf{X}^\top
\end{equation}
is built from the matrix $\mathbf{X}$ whose columns are the powers $\nu^n$ evaluated each channel. We note that $\mathbf{M}$ depends only on the observing frequencies and the polynomial order, and not on the data itself. The same filter therefore removes any other spectrally smooth component present in the data. In particular, because the CMB is frequency independent in thermodynamic units, it has the same smooth spectral shape in every pixel and is therefore almost totally absorbed by the quadratic fit. Figure~\ref{fig:cib-clean} confirms this on mocks with and without a CMB component. Both the filtered spectra and their error bars are unchanged at the $1\%$ level.

As $\mathbf{M}$ is a linear operator, its effect on the signal power spectrum can be computed analytically as in \cite{Liu2011}. Given the full matrix of cross-frequency signal covariances $\mathbf{C}_\ell$, the cleaned covariance is
\begin{equation}
    \label{eq:mcm}
    \mathbf{C}_\ell^\mathrm{clean} = \mathbf{M} \,\mathbf{C}_\ell \,\mathbf{M}^\top \ ,
\end{equation}
and so the transfer function applied to each filtered channel pair $(a,b)$ is the ratio of cleaned to input cross-power:
\begin{equation}
    F_\ell^{ab} = \frac{\left(\mathbf{M} \mathbf{C}_\ell \mathbf{M}^\top \right)_{ab}}{\left(\mathbf{C}_\ell\right)_{ab}} \ .
\end{equation}
By design, this transfer function carries no additional noise forward to the parameter inference. However, as in the case of the likelihood covariance, we calibrate the signal covariance $\mathbf{C}_\ell$ from line-only mock cubes produced by SIDES. For the $g\nu$ cross-spectra only the LIM cube is filtered, and we mask the expected line emission before the polynomial fit to avoid absorbing the signal (Appendix~\ref{sec:photo-z-masking}).

\begin{figure}
    \centering
    \includegraphics[width=\linewidth]{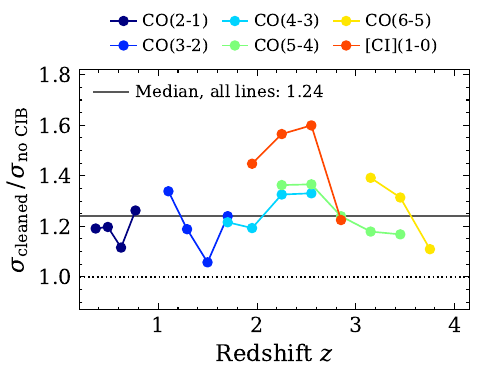}
    \caption{The cost of continuum removal on the line amplitude constraints, shown as the ratio of marginalized error on one bias-weighted intensity $\langle bI\rangle_J(z_b)$ between the nominal cleaned fit and an otherwise identical fit to a CIB-free cube where no foreground cleaning is needed. Values above unity indicate that the cleaning procedure costs precision to parameter inference. The grey line and band show the mean over all lines and redshift bins, $1.28\pm0.06$. We stress that the continuum removal adopted here has not been optimized, and that a more refined cleaning procedure should reduce the impact to parameter constraints.
    }
    \label{fig:cib-cost}
\end{figure}

Although this correction restores the mean line power in the signal, the cleaning itself is not free. As is evident in Figure \ref{fig:cib-clean}, the polynomial removes modes that cannot be recovered, inflating the error bars in the data and ultimately in the inferred constraints. Figure~\ref{fig:cib-cost} quantifies this cost at the parameter level, comparing the marginalized error on each line amplitude between the nominal cleaned fit and an otherwise identical fit to a continuum-free cube where no foreground cleaning is needed. Cleaning degrades the amplitude constraints by an average of about $30\%$ across the redshift range $0.3$ to $3.0$ considered for this configuration.

\section{Joint Likelihood}
\label{sec:like}
We construct a Gaussian likelihood for the joint data vector $\mathbf{D}$ combining the four available blocks of auto- and cross-spectra:
\begin{equation}
    \mathbf{D} \;=\; \{\,C_\ell^{\nu\nu},\; C_\ell^{\nu\nu'},\; C_\ell^{g\nu},\; C_\ell^{gg}\,\} \ .
\end{equation}
We assume the standard form
\begin{equation}
\begin{split}
    \mathcal{L}(\boldsymbol{\theta} \mid \mathbf{D}) \propto \exp\bigg[
    & -\frac{1}{2} \sum_\ell \big(\mathbf{D}_\ell - \mathbf{M}_\ell(\boldsymbol{\theta})\big)^{\!\top} \mathbf{C}_\ell^{-1} \\
    & \times \big(\mathbf{D}_\ell - \mathbf{M}_\ell(\boldsymbol{\theta})\big) \bigg] \ ,
\end{split}
\end{equation}
where $\mathbf{M}_\ell(\mathbf{\theta})$ is the model vector of angular power spectra evaluated at parameters $\mathbf{\theta}$ and $\mathbf{C}_\ell$ is the band-power covariance matrix.

The data vector $\mathbf{D}$ includes all measured band powers across the four blocks. For the LIM block with $N_\nu$ frequency channels, this consists of $N_\nu$ auto-spectra plus every unique cross-frequency pair $(\nu_i,\nu_j)$, for a total of $N_\nu(N_\nu+1)/2$ power spectra. For $gg$ it is every unique pair of galaxy bins and for $g\nu$ it is the product of galaxy bins with LIM channels. Data points which are expected to be uncorrelated, i.e. $C_\ell^{ab}$ where $a$ and $b$ contain no tracers that overlap in redshift, are discarded from the data vector. The covariance is then estimated over the retained band powers only, which substantially reduces the computational sampling time.

For our nominal LIM~$\times$~photo-$z$ configuration with 51 LIM channels spanning $125 \ \mathrm{GHz}<\nu<175 \ \mathrm{GHz}$ and 9 photo-$z$ galaxy bins, this amounts to 249 non-zero band powers per $\ell$ bin: 51 $\nu\nu$, 95 $\nu\nu'$, 94 $g\nu$, and 9 $gg$ spectra. Replacing the photo-$z$ catalog with a spectroscopic galaxy catalog covering approximately the same redshift range (see Appendix \ref{sec:spec-z}) further grows the data vector, since the finer redshift binning produces more $gg$ and $g\nu$ pairs. Despite the large data vector size, the joint likelihood remains computationally modest since each band power is estimated from Fourier transforms of the input maps and catalogs in the flat sky approximation (Section \ref{sec:FT}). A more rigorous estimation of the power spectrum that incorporates spherical harmonic transforms and mode-coupling matrices, e.g. \texttt{NaMaster}~\citep{Alonso2019}, would be substantially more expensive at this dimensionality. However, this only becomes necessary for larger sky areas or when realistic survey masks introduce non-trivial mode coupling between the LIM and galaxy footprints.

\subsection{Data Templates}
\label{sec:templates}
\begin{figure}
    \centering
    \includegraphics[width=\linewidth]{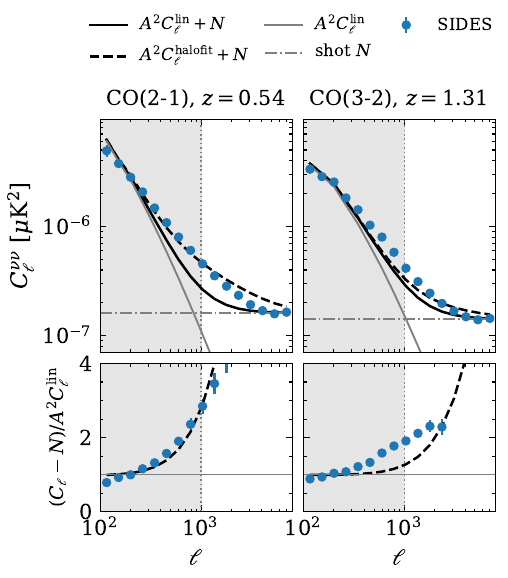}
    \caption{\textit{Top:} Auto-spectra of two different lines in a 150~GHz channel, measured from ideal mocks with no interlopers or foregrounds. Points show the mean over the 16 mock realizations. The solid and dashed black curves show the linear and halofit Limber predictions plus a flat shot noise term $N$. The shaded region marks the scales used in this analysis, $100<\ell<1000$. \textit{Bottom:} Clustering power relative to the linear prediction, where values above 1 indicate excess clustering power not captured by a standard linear template. At low redshift the excess is captured by the halofit model, while at higher redshift it exceeds the non-linear matter prediction.}
    \label{fig:templates}
\end{figure}

Each band power in the model vector $\mathbf{M}_\ell(\mathbf{\theta})$ is a linear combination of angular clustering templates scaled by the free amplitude parameters,
\begin{equation}
    M_\ell^{ab} = A_a\,A_b\,S_\ell^{ab} + \rho_{ab}\sqrt{N_a N_b}
\end{equation}
where $A_a$ is the amplitude of field $a$ ($\langle bI\rangle$ for a line or $b_g$ for a galaxy bin), $S_\ell^{ab}$ is the angular clustering template for the pair $(a,b)$, and $N_a$ is the respective shot noise term. Templates are constructed empirically from the mock suite. First, we recover a set of fiducial parameters that encode the true amplitudes of the mock signal by fitting each line and galaxy bin using ideal mocks with no noise, interlopers or foregrounds. Each template is then the ensemble averaged power spectrum of the pair $(a,b)$, with the fiducial amplitudes divided out and fiducial shot noise subtracted. This ensures that the model recovers the mean mock data when evaluated at the fiducial amplitudes, so the recovered $A_a$ are measured relative to this normalization. This empirical approach captures the true $\ell$-dependence of each signal without relying on an analytic Limber computation, which could otherwise miss non-linear physics and survey systematics present in the mocks. Figure~\ref{fig:templates} shows that this non-linear contribution is non-negligible over the scales used in this analysis, particularly at higher redshifts where it also exceeds the halofit prediction. We note that $A_a$ is therefore strictly the amplitude of this empirical clustering shape, and coincides with $\langle bI \rangle$ only in the linear regime. A halo model template that separately parameterizes the 2-halo and 1-halo contributions would recover $\langle bI\rangle$ directly, which we leave to future work. In Section \ref{sec:robust}, we assess how imperfect template knowledge would propagate to inferred parameters.

\subsection{Covariance Matrix}
\label{sec:covariance}
We estimate the covariance between all pairs of band powers in the data vector using the analytic form as in e.g. \cite{Hu2004} and \cite{Krause2017}:

\begin{equation}
\mathrm{Cov}\!\left(C_\ell^{ab}, C_\ell^{cd}\right) = \frac{1}{(2\ell+1)\,\Delta\ell\, f_\mathrm{sky}} \left[ C_\ell^{ac} C_\ell^{bd} + C_\ell^{ad} C_\ell^{bc} \right] \ ,
\label{eq:knox}
\end{equation}
where $a, b, c, d$ index the fields in the joint data vector and $f_\mathrm{sky}$ is set by the effective survey area of Table~\ref{tab:config}. The products on the right-hand side are evaluated at the fiducial parameter values obtained from the mock data. 

This form considers only Gaussian contributions to the covariance. Strictly speaking, this is an approximation even at the scales considered in this work ($\ell<1000$). Non-Gaussian contributions arise from non-linear clustering and shot noise, and grow at smaller scales. For tracers with broad redshift kernels, such as photo-$z$ galaxy samples, these contributions are suppressed by averaging over many independent structures along the line of sight. However, because each LIM channel probes a narrow redshift slice, the suppression here is much weaker. These contributions are expected to primarily affect the size of the error bars rather than bias the recovered amplitudes. A more complete treatment would estimate the covariance either using a halo model formalism or from an ensemble of simulations. The primary limitation of the latter approach is that a robust inversion of the covariance matrix requires $N_\mathrm{sim}\gg N_\mathrm{data}$ \citep{Taylor2013}. For our data vector with $N_\mathrm{data}\sim250$ per $\ell$ bin, this implies several thousand simulated realizations.

\begin{figure*}
    \centering
    \includegraphics[width=0.8\linewidth]{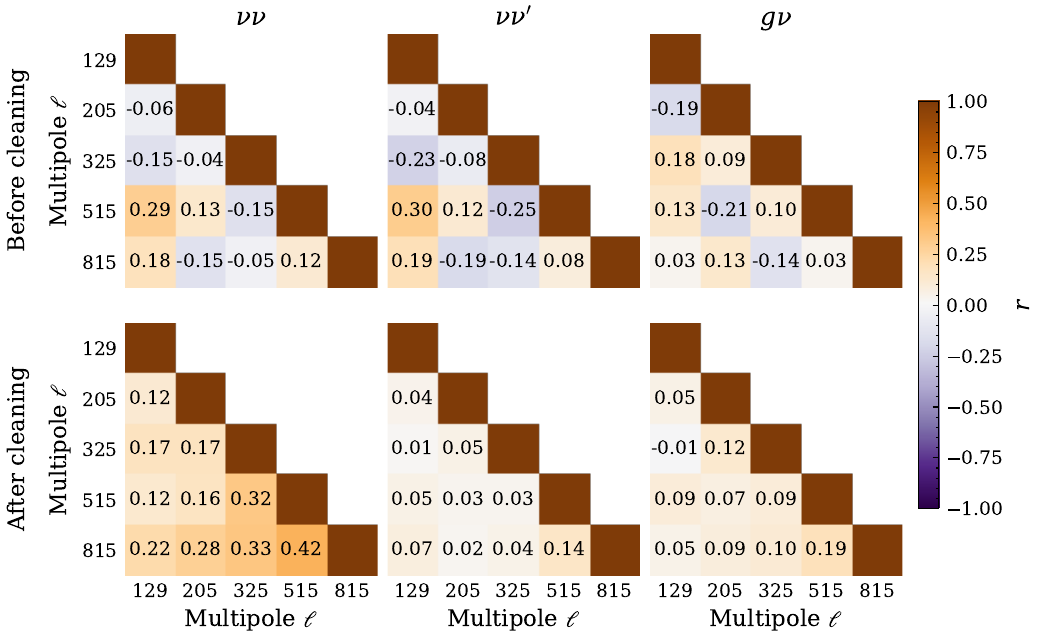}
    \caption{Band power correlations measured across 16 mock tiles, before (top row) and after (bottom row) applying the CIB continuum filter. Values shown are the median over all spectra of a given type, i.e. line auto ($\nu \nu$), line-line cross ($\nu \nu'$), and galaxy-cross ($g \nu$). Prior to cleaning, the off-diagonal correlations are dictated by realization noise and are consistent with zero everywhere. After cleaning, the auto-spectra develop a positive correlation that grows with $\ell$, up to $r\approx0.4$ between the two highest pairs at $\ell=515$ and $\ell=815$. The two cross-spectra blocks show much smaller off-diagonal correlations in comparison.
    }
    \label{fig:bandpower-corr}
\end{figure*}

Equation~\ref{eq:knox} is also diagonal in $\ell$ by construction. The continuum filter described in Section~\ref{sec:cib} leaves a residual whose amplitude varies between realizations and scales coherently across the band, which correlates band powers that the analytic form treats as independent. Figure~\ref{fig:bandpower-corr} shows the band power correlation measured across 16 mock tiles, before and after foreground cleaning. Before cleaning, the off-diagonal correlations are consistent with zero in every data block. After cleaning, the line auto-spectra $\nu \nu$ develop a positive correlation that grows with scale, reaching $r\approx0.4$ between the two highest $\ell$ band powers, while the line-cross and galaxy-cross blocks remain largely unaffected.

\subsection{Parameterization}
We adopt a model agnostic parameterization in which line signals are binned directly in redshift space, with no assumptions on line ratios or redshift evolution. To define these parameters, we divide the native LIM frequency space into a redshift grid with cell bins $z_b$ and widths $dz_b\sim0.3$. The model vector is parameterized by the bias-weighted mean intensity of each spectral line in every redshift bin, $\langle bI\rangle_{J_i}(z_b)$, and effective shot noise amplitudes per $z$-bin $N_{J_i}(z_b)$. We additionally fit a linear galaxy bias $b_g(z_b)$ for every galaxy bin $z_b$. The galaxy shot noise is not fit but fixed to the catalog Poisson value $N_g=1/\bar{n}$, since the 1-halo clustering contribution is already accounted for in the $gg$ template. The number of $\langle bI\rangle$ parameters is set by the number of grid cells in which each line's emission falls within the observable LIM frequency range. For the baseline LIM~$\times$~photo-$z$ configuration, this amounts to 45 bias-weighted intensity amplitudes and 45 line shot amplitudes (one of each per line per occupied $z$-bin), plus 9 galaxy bias parameters and 8 shot correlation coefficients $\rho_{ij}$, for a total of 107 sampled parameters. For all amplitudes, we place broad log-uniform priors. The full parameter set is summarized in Table~\ref{tab:params}.

\begin{deluxetable}{lllc}
  \tablecaption{Parameters in the joint likelihood for the baseline
  LIM~$\times$~photo-$z$ configuration. Counts assume 7 CO transitions plus the [CI](1--0) line covered
  over $0.3<z<6.5$ with $dz_b\sim0.3$ and a photo-$z$ galaxy sample contributing 9 bias parameters.\label{tab:params}}
  \tablehead{
    \colhead{Parameter} &
    \colhead{Description} &
    \colhead{Prior} &
    \colhead{Count}
  }
  \startdata
    $\langle bI\rangle_{J_i}(z_b)$ &
      Bias-weighted line intensity per line per $z$-bin &
      $[10^{-2},\,10^2]$ &
      45\\
    $N_{J_i}(z_b)$ &
      Line shot noise per line per $z$-bin &
      $[10^{-10},\,10^{-5}]$ &
      45 \\
    $b_g(z_b)$ &
      Linear galaxy bias per photo-$z$ bin &
      $[0.5,\,5.0]$ &
      9 \\
    $N_g(z_b)$ &
      Galaxy shot noise per photo-$z$ bin &
      fixed at $1/\bar n$ &
      --- \\
    $\rho_{ij}$ &
      Shot noise correlation coefficient per line pair &
      $[0,\,1]$ &
      8 \\
  \enddata
\end{deluxetable}

\section{Results}
We apply the joint likelihood framework of Section \ref{sec:like} to the mock datasets described in Section~\ref{sec:mocks} to assess the extent to which multi-tracer cross-correlations can separate overlapping CO lines. Our baseline mock, summarized in Table~\ref{tab:config}, consists of a LIM survey covering $125<\nu<175$~GHz (SPT-SLIM-like) over a $100~\mathrm{deg}^2$ patch, cross-correlated with an LSST Y1-like photometric galaxy catalog spanning $0.2<z<1.6$ with scatter $\sigma_z = 0.03(1+z)$. The continuum foregrounds (CIB and CMB) are removed from the dataset using the line-of-sight polynomial fitting procedure as described in Section~\ref{sec:cib}. Within the 125--175~GHz band, this configuration is sensitive to 7 CO rotational transitions, CO(2-1) through CO(8-7), as well as [CI](1-0). Together, these lines span a combined redshift range of $0.3\lesssim z\lesssim 6.5$ over the SPT-SLIM frequency window. Although [CII] and higher-order CO transitions are present in the mock dataset, these signals are significantly fainter and are considered negligible for this analysis.

\begin{figure*}
    \centering
    \includegraphics[width=0.9\linewidth]{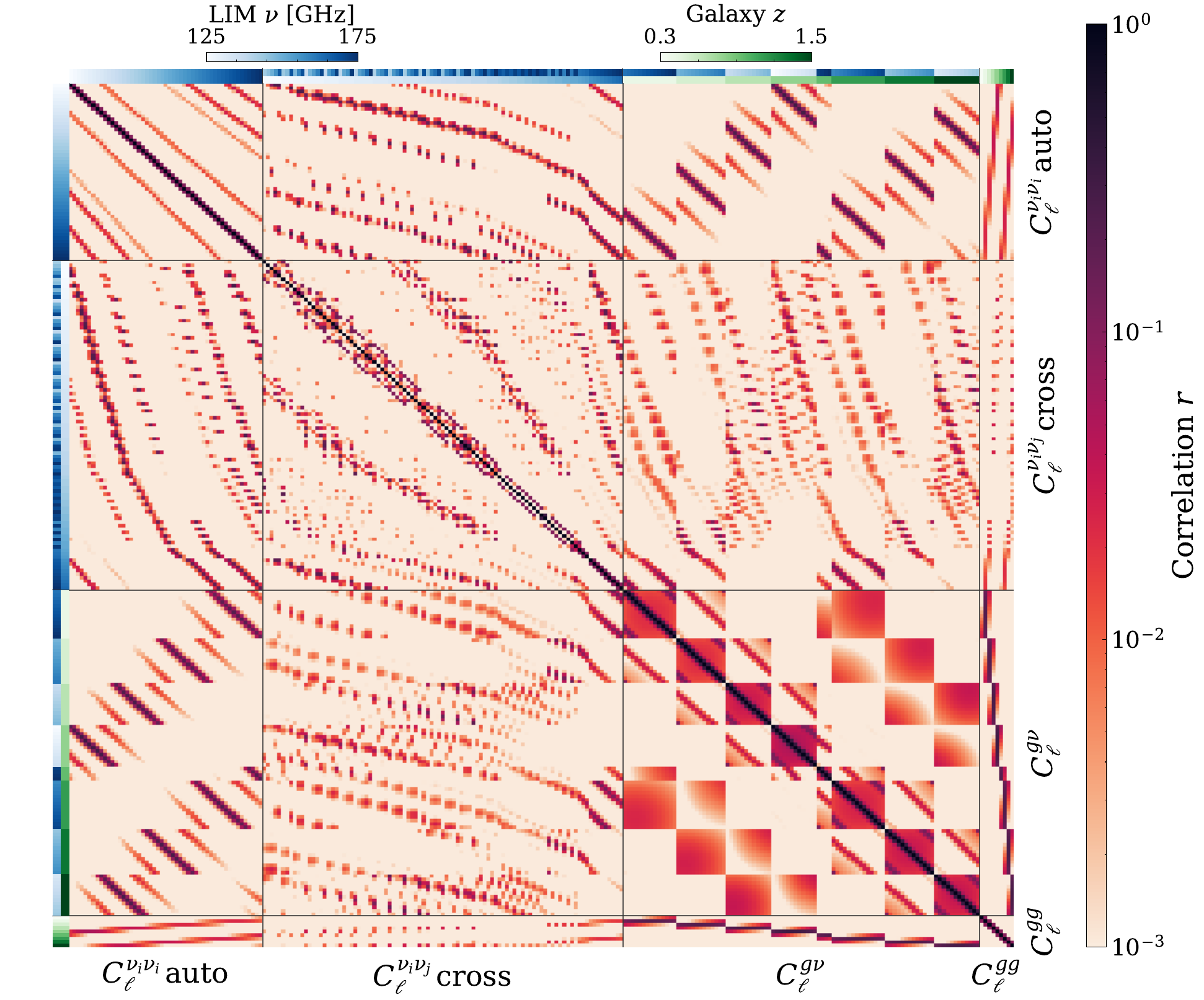}
    \caption{Correlation matrix of the joint data vector $\mathbf{D}=\{C_\ell^{\nu\nu},\,C_\ell^{\nu\nu'},\,C_\ell^{g\nu},\,C_\ell^{gg}\}$ in a single $\ell$ bin, obtained from the covariance described in Section~\ref{sec:covariance}. Each row and column corresponds to one band power rather than one frequency, and the colored strips along each axis identify the pair of fields that make up each band power. Off-diagonal correlations appear wherever two band powers share the same line emission. In the line blocks, these connect channels that are either adjacent in frequency or whose frequency ratio matches a line ratio so that both sample the same redshift. The $g\nu$ block shows broader Gaussian-shaped correlations set by the width of the photo-$z$ kernels, while the galaxy auto-spectra are nearly independent.}
    \label{fig:covmat}
\end{figure*}

To isolate the contribution of the galaxy tracer to interloper separation, we compare two configurations. The LIM-only fit uses the auto- and cross-spectra internal to the LIM dataset, $\mathbf{D}_\mathrm{LIM} = \{C_\ell^{\nu\nu}, C_\ell^{\nu\nu'}\}$. The joint LIM + photo-$z$ fit adds the galaxy-line and the galaxy-galaxy spectra, $\mathbf{D} = \{C_\ell^{\nu\nu}, C_\ell^{\nu\nu'},C_\ell^{g\nu}, C_\ell^{gg}\}$, whose correlation matrix is shown in Figure~\ref{fig:covmat}. Chains are run in \texttt{cobaya} with a standard MCMC sampler.

\subsection{Constraints from LIM-only}
\begin{figure*}
    \centering
    \includegraphics[width=\linewidth]{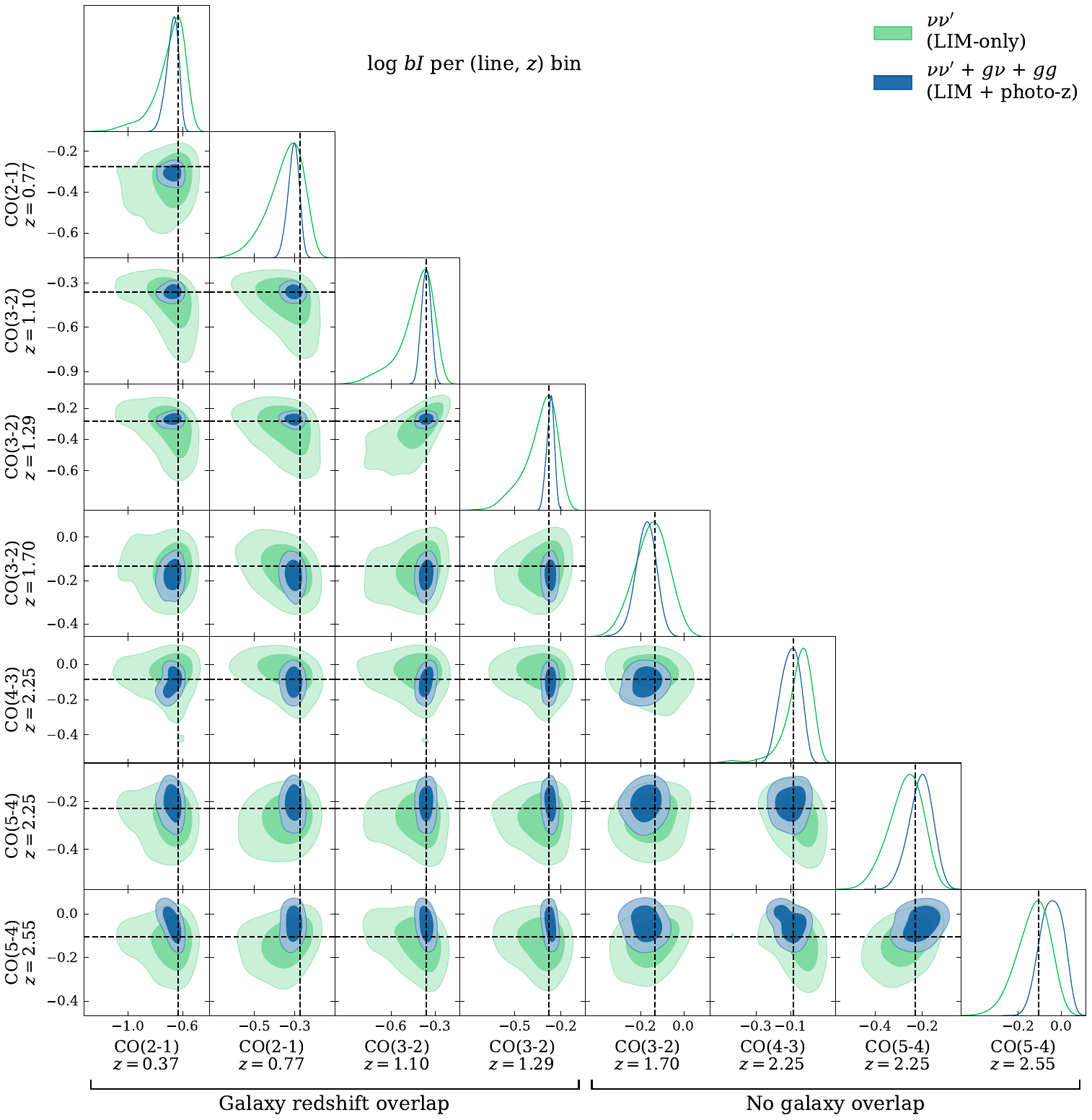}
    \caption{1D and 2D marginalized posteriors on $\log\langle bI\rangle$ for a representative subset of $(J, z)$ bins observable in the SLIMxLSST dataset, split between bins inside the LSST photo-$z$ range ($z \lesssim1.6$) and beyond direct galaxy overlap. Green contours use the LIM-only configuration, while blue contours use the full LIM + photo-$z$ joint fit. Filled contours show the $1\sigma$ and $2\sigma$ uncertainties, and dashed lines indicate the fiducial truth. At $z<1.6$, where the galaxy sample overlaps the lines directly, adding galaxy data reduces the marginalized errors by a factor of $\sim$3--4. At higher redshift, the errors still shrink by a factor of $\sim$1.2--1.5, since the galaxy data constrain the low-$z$ interlopers that share frequency channels with these lines.}
    \label{fig:corner}
\end{figure*}

Figure \ref{fig:corner} shows marginalized posteriors on a representative subset of $\log\langle bI\rangle$ parameters for the LIM-only configuration (green contours). Even with the LIM cross-spectra included, degeneracies continue to persist across the full range of observables, particularly at low redshifts where line-line cross-correlations are sparse. These posteriors reflect the fundamental challenge of interloper separation. A single frequency channel at $\nu\sim150$~GHz will receive comparable signal power from CO(2-1) at $z\sim0.5$, CO(3-2) at $z\sim1.3$, CO(4-3) at $z\sim2.1$, and CO(5-4) at $z\sim2.8$, and the auto-spectrum $C_\ell^{\nu\nu}$ alone cannot assign the observed power between them. The LL cross-spectra can partially break this degeneracy by isolating pairs of lines that share a common redshift, but the number of such accessible same-$z$ pairs within a 50~GHz frequency range is limited.

\subsection{Constraints from LIM + Galaxy}
\label{sec:joint_results}
Including the galaxy data $gg+gl$ into the likelihood produces qualitatively different posteriors, as shown by the blue contours in Figure~\ref{fig:corner}. We note that many of the degeneracies seen in the LIM-only fit collapse significantly once the galaxy sample is included in the data vector. The improvement is driven by the tomographic information carried by the photo-$z$ bins, which anchor the otherwise-unconstrained low-$z$ lines and alleviate degeneracies in the LIM auto spectra. The posterior on the photo-$z$ galaxy bias $b_g(z_g)$ is simultaneously constrained by the galaxy auto-spectrum $C_\ell^{gg}$, which breaks the $\langle bI\rangle$--$b_g$ degeneracy that would otherwise persist in Eq.~\ref{eq:glcross}.

\begin{figure}
    \centering
    \includegraphics[width=\linewidth]{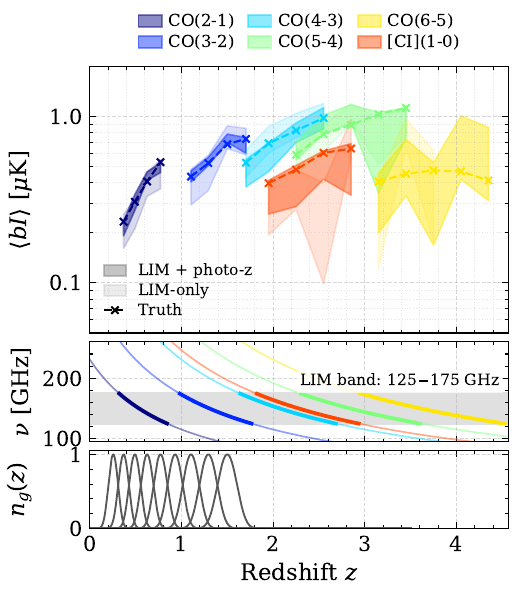}
    \caption{Recovery of the bias-weighted mean intensity, $\langle bI\rangle(z)$, for all significant emission lines observable in the 125-175~GHz frequency window (top panel). Shaded regions show the $1\sigma$ intervals per $z$-bin, with darker (lighter) shading for the joint LIM + photo-$z$ (LIM-only) analysis. Dashed curves represent the fiducial truth from the simulation data. The middle panel shows the observed frequency $\nu_\mathrm{obs} = \nu_\mathrm{rest}/(1+z)$ of each transition, with the segment falling inside the survey band (grey) highlighted: each line is constrained exactly over the redshift window in which it redshifts into the narrow band. The bottom panel shows the normalized redshift distributions of the photometric galaxy bins. Constraints hold across the full redshift range accessible to each transition, including bins above the galaxy-survey horizon $z_\mathrm{max}^{g}=1.6$.}
    \label{fig:tomography}
\end{figure}

Figure~\ref{fig:tomography} shows these results in terms of the redshift evolution of the bias-weighted mean intensity $\langle bI\rangle(z)$ for each CO transition. The shaded bands show the $1\sigma$ interval per $z$-bin alongside the fiducial truth. Two features are worth highlighting in this figure. First, within the redshift coverage of the galaxy survey ($z<1.6$), the recovered amplitudes are tightly constrained and unbiased with respect to the fiducial, since the GL cross-spectrum is directly anchoring the fit here. Second, and less obvious, is that the addition of galaxy data also yields moderately better constraints on the high-$z$ lines that lie beyond the galaxy survey coverage. By tightly constraining the low-$z$ interlopers through the $g\nu$ cross-spectra, the galaxy data effectively clean up the auto-power spectra, leaving the high-$z$ line signals to be fit by the LIM-only block of the data vector. We expect this effect to be even more significant for lines with no available cross-spectra of their own, e.g., [CII] observable in the higher-frequency windows (see Section \ref{sec:cii}).

\begin{figure}
    \centering
    \includegraphics[width=\linewidth]{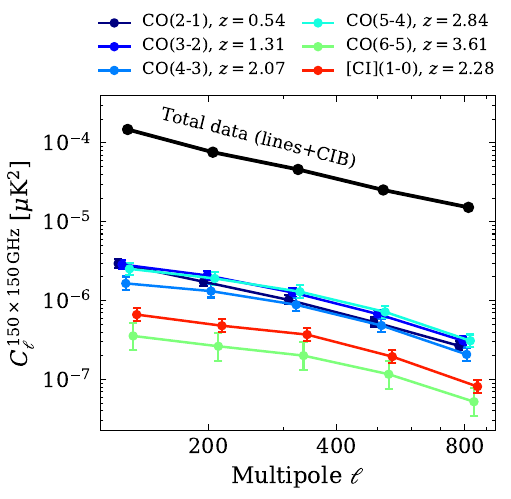}
    \caption{Spectral decomposition of a single LIM auto-frequency channel at $\nu=150$~GHz. Colored points show the per-line clustering contributions to $C_\ell^{\nu\nu}$, reconstructed from the amplitudes recovered in the joint LIM + photo-$z$ fit, with error bars showing the $1\sigma$ posterior uncertainties. Each line contributes from a distinct redshift, indicated in the legend. The black curve shows the total uncleaned power in the channel, which is dominated by the CIB continuum.}
    \label{fig:decomp}
\end{figure}

The component separation enabled by the joint fit is summarized in Figure~\ref{fig:decomp}, which decomposes the observed auto-power of a single LIM channel at $\nu=150$~GHz into the recovered contribution of each spectral line. At this frequency, the recovered amplitudes assign the observed power to individual transitions spanning CO(2-1) at $z=0.5$ through CO(6-5) at $z=3.6$, with each contribution lying orders of magnitude below the CIB-dominated total signal. This figure illustrates the full extent of the component separation problem addressed in this work, where each line contribution must be recovered in the presence of both the bright continuum foreground and the many overlapping lines, none of which are individually accessible from the auto-spectrum alone.

\subsection{Model Robustness}
\label{sec:robust}
\begin{figure}
    \centering
    \includegraphics[width=\linewidth]{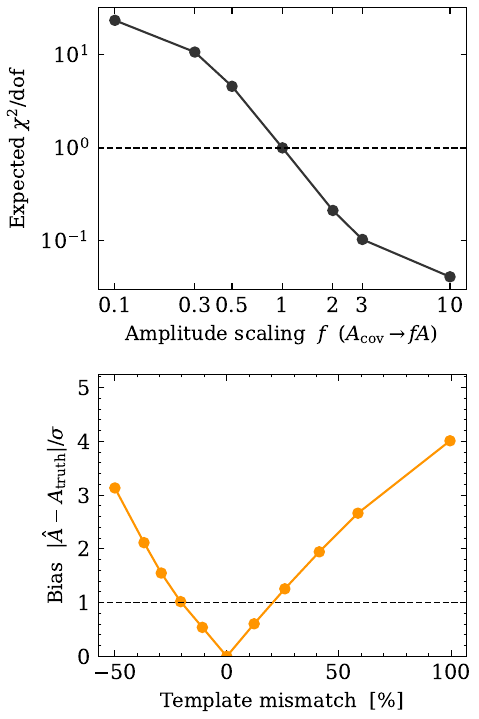}
    \caption{Robustness of the recovered line amplitudes to model misspecification. \textit{Top}: reduced $\chi^2$ as a function of the rescaling factor $f$ applied to the fiducial line amplitudes assumed in the covariance. The recovered amplitudes remain unbiased for all $f$, with the wrong amplitudes impacting only the goodness of fit, with $\chi^2/\mathrm{dof}=1$ at the ground truth $f=1$. \textit{Bottom}: maximum bias on the recovered amplitudes, in units of $\sigma$, as a function of the template mismatch, i.e. the fractional distortion of the template shape across $100<\ell<1000$. Shape errors bias parameters at the $1\sigma$ level when the template mismatch approaches $\sim25\%$.}
    \label{fig:robustness}
\end{figure}

The robustness of these constraints depends primarily on two model inputs: the fiducial line amplitudes assumed in the covariance (Section~\ref{sec:covariance}), and the power spectrum shapes assumed in the templates (Section~\ref{sec:templates}). Figure~\ref{fig:robustness} (top) shows the impact of mis-estimating the clustering amplitudes on the reduced $\chi^2$ statistic, which we evaluate by rescaling the amplitudes $A$ of every line in the covariance by a factor $f$, up to an order of magnitude in both directions to reflect the current large uncertainties on these parameters. Since the fiducial amplitudes only enter in the covariance, the recovered parameters remain unbiased regardless of the inaccuracy in the initial amplitude guess. Instead, the line amplitude mis-specification results in a predictable impact on the reduced $\chi^2$, with $\chi^2/\mathrm{dof}=1$ at the ground truth $f=1$.

Template shape errors, on the other hand, propagate directly into the recovered parameters. We test this by tilting the angular shape of every template by a power law, $S_\ell^{ab} \rightarrow S_\ell^{ab}\,(\ell/\ell_\mathrm{piv})^\epsilon$, anchored at a pivot scale $\ell\sim400$ such that the mean amplitude is preserved and $\epsilon$ acts as a pure shape error. We quantify this as the template mismatch, the fractional distortion of the template shape across our full multipole range $100<\ell<1000$. Figure~\ref{fig:robustness} (bottom) shows the resulting bias on the recovered amplitudes as a function of the template mismatch percentage. The recovered amplitudes become biased at the $1\sigma$ level once the template mismatch becomes greater than $\sim25\%$, corresponding to $|\epsilon|\approx0.1$. This sets the accuracy to which the clustering shapes must be modeled for the amplitude recovery to remain unbiased. This requirement is most relevant for the nonlinear clustering at $\ell\gtrsim400$ (Figure~\ref{fig:templates}), which at higher redshifts is driven by the 1-halo term or scale-dependent bias of the line emitters rather than by the matter power spectrum. Since this contribution would not scale with $\langle bI\rangle$ in the same way as the 2-halo term, its shape will in general differ between simulations and the true sky, and templates for real data will need to model it to this level of accuracy.

\subsection{[CII] recovery}
\label{sec:cii}
At the frequencies considered in this work (125-175 GHz), [CII] is a negligible contribution to the observed sky power. However, this multi-tracer approach could be straightforwardly applied to higher frequencies to recover [CII] emission from $z\approx5-10$, as in spectral bands targeted by e.g., CONCERTO, TIME, and CCAT. In this regime, direct cross-correlations with the [CII] signal are largely unavailable, since the line lies well beyond the redshift coverage of galaxy surveys and no other bright transitions from the same redshift are observable at these frequencies. However, as we have shown in this work, the lower redshift CO and [CI] interlopers that contaminate the [CII] channels can be well-constrained by the galaxy and line-line cross-correlations. Following the same mechanism discussed in Section~\ref{sec:joint_results}, constraining these interlopers effectively cleans up the [CII] auto-power spectra, allowing the [CII] line to be recovered despite having no direct cross-correlations of their own. We leave the optimization of high-redshift [CII] line deconfusion to future work.

\section{Conclusion}
In this work, we have demonstrated the effectiveness of a joint likelihood approach incorporating multiple sources of information to simultaneously recover individual lines from a realistic line intensity mapping dataset. While the impact of line interlopers can be mitigated by masking or discarding contaminated data, we show that an optimal 2-point analysis of LIM data in the high line confusion regime requires an additional probe, such as internal or external cross-correlations, to break the degeneracy intrinsic to the auto-correlation measurement. 

Moreover, we show that the multi-tracer approach to line separation offers constraints that are more than the sum of its parts. By combining LIM-galaxy cross-correlations that anchor the low-redshift lines with LIM-LIM cross-correlations that constrain the high-redshift lines, we are able to utilize the full redshift range available in the data. This is especially critical for experiments with limited frequency range, where few transitions from the same redshift fall simultaneously within the observing band and the number of available internal cross-correlations is limited. These insights point to the simple reality that future line intensity mapping surveys must consider the availability of both internal and external cross-correlations as a primary driver of component separation. 

Although we have only considered a joint likelihood incorporating two tracers, LIM and photo-$z$ galaxy surveys, this method can be generalized to any number of LSS tracers provided that the data covariance can be reasonably estimated. Additionally, we note that this framework is not limited to two-point statistics. For example, \cite{Ihle2019} showed that including the one-point voxel intensity distribution (VID) in a joint analysis with the power spectrum can significantly enhance signal recovery in an intensity mapping survey. Furthermore, although we do not attempt to break the bias-intensity degeneracy in the recovered line amplitudes in this work, a joint analysis of the VID + multi-tracer cross-spectra could be an effective approach to recovering separate bias and mean intensity in the regime of many line interlopers \citep{Breysse2022}.

A key aspect of this work is evaluating how multi-line recovery is affected by the CIB continuum foreground and its inherent spatial correlation with the target emission lines. Although we demonstrate that a pixel-level polynomial fit can effectively remove the CIB foreground, the associated line signal loss results in a non-negligible impact to parameter constraints (Figure \ref{fig:cib-cost}). This motivates a more careful optimization of foreground cleaning than what is done here. Several directions that could improve the foreground removal include a different functional form that is better matched to the continuum signal, spectral parameters that are correlated across the sky rather than fitted independently along each line of sight, or the use of spatial templates. A more ambitious approach would be to model the CIB foreground as an additional component in the likelihood, with its own spectral energy distribution and correlations to each line and galaxy bin. We leave a systematic comparison of these approaches to future work.

A related question is how much the degradation shown in Figure~\ref{fig:cib-cost} depends on the particular line model used to generate the mocks. Line power removed by a polynomial filter depends on how the brightness of the line evolves with redshift, which is set by the line's luminosity density and excitation ladder. Repeating this comparison across mocks spanning the current uncertainty in CO, [CI] and [CII] modeling would establish whether the $\sim30\%$ constraint degradation we quote here is a robust estimate or specific to the simulations we adopt.

These results also rely on clustering templates measured directly from the mock ensemble. While this ensures a self-consistent test of our joint framework, the templates include non-linear clustering contributions that may not match the true sky signal. Applying this framework to real data will therefore require templates that model these contributions directly, e.g. with a halo model calibrated against simulations and, eventually, constrained by the data themselves.

Looking ahead, experiments such as TIME, SPT-SLIM, CCAT, and CONCERTO are positioned to soon deliver the first high-significance detections of the CO and [CII] clustering signals at millimeter wavelengths, likely in combination with overlapping LSS surveys such as LSST, DESI, SPHEREx, and \emph{Euclid}. The multi-tracer framework developed here is designed precisely for this regime, and we argue that the availability of such overlapping datasets should be treated as a primary consideration in the design of future LIM surveys in order to extract the full cosmological information encoded in these fields. 

\begin{acknowledgments}
We thank Emmanuel Schaan, José Luis Bernal, Yun-Ting Cheng, Jamie Bock, Jessica Zebrowski, Zhuoqi (Jackie) Zhang, and Zeeshan Ahmed for their very helpful comments.

CS was supported by the Department of Energy, Laboratory Directed Research and Development program at SLAC National Accelerator Laboratory, under contract DE-AC02-76SF00515.

\end{acknowledgments}

\appendix
\section{Line-line and galaxy-line decorrelation}
\label{sec:decorrelation}
The formalism in Section~2 assumes perfect correlation between the clustering of any two tracers at the same redshift, $r_{ij}=1$. We test this assumption in the ideal mocks by measuring the correlation coefficient
\begin{equation}
    r(\ell)=\frac{C_\ell^{ab}}{\sqrt{C_\ell^{aa}\,C_\ell^{bb}}}
\end{equation}
for a set of line-line and galaxy-line pairs. Since each field is a projection over its own redshift kernel, two perfectly correlated fields will still have $r<1$ whenever their kernels differ in width. We correct for this geometric effect by dividing the coefficient expected from the kernels alone.

\begin{figure}
    \centering
    \includegraphics[width=0.8\linewidth]{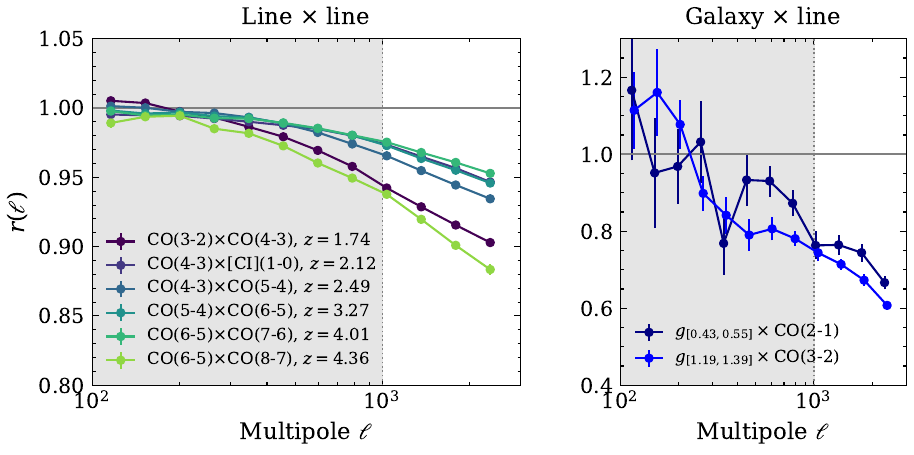}
    \caption{Correlation coefficient between line pairs (left) and between lines and photo-$z$ galaxy bins (right) at the same redshift, measured from ideal mocks and corrected for projection effects. The shaded region marks the scales used in this analysis.}
    \label{fig:decorrelation}
\end{figure}

Figure ~\ref{fig:decorrelation} shows that line pairs at the same redshift remain correlated to within $\sim5\%$ of unity out to $\ell=1000$. The gradual decline at high $\ell$ is driven by the shot noise term, whose decorrelation $\rho_{ij}<1$ is expected at these scales as we are measuring the scatter in line ratios between individual galaxies. For this reason we fit for these shot noise decorrelation terms in the likelihood. The galaxy-line pairs are consistent with unity at low $\ell$, and fall to $r\approx0.75$ by $\ell=1000$. This decline reflects the decorrelation between galaxy and line shot noise, which is already captured by the model since the $g\nu$ templates in our analysis are measured directly from the mock ensemble (Section~\ref{sec:templates}).

\section{Comparing photo-$z$ and spec-z constraints}
\label{sec:spec-z}
\begin{figure}
    \centering
    \includegraphics[width=0.5\linewidth]{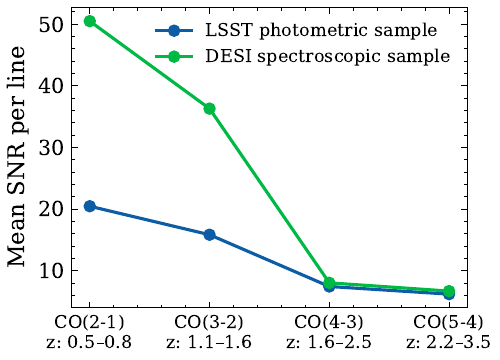}
    \caption{Forecast signal-to-noise ratio of the CO line amplitudes in each redshift bin for the SPT-SLIM-like 100 $\mathrm{deg}^2$ survey, using LSST photometric (blue) and DESI spectroscopic (green) galaxy samples. The spectroscopic sample tightens the constraints by 2-3x where it overlaps the line in redshift ($z \lesssim 1.6$), but only marginally at higher redshift.}
    \label{fig:snr}
\end{figure}

Figure~\ref{fig:snr} compares the Fisher forecast signal-to-noise ratio on the line amplitudes, averaged over $z$-bins, for the joint fit with an LSST-like photometric sample ($\sigma_z=0.03(1+z)$) and with a DESI-like spectroscopic sample ($\sigma_z=0$) that is significantly shallower over the same redshift range. For the low-$J$ lines with direct galaxy redshift overlap, the spectroscopic sample improves the signal-to-noise ratio by a factor of 2--3. For the higher-$J$ lines, the gains from the spectroscopic sample are a much more modest $\sim$10\%.

\section{Filtering considerations for LIM x photo-$z$}
\label{sec:photo-z-masking}
The foreground filtering operation of the LIM data cubes described in Section \ref{sec:cib} can introduce significant loss to the cross-correlation with photo-$z$ galaxy catalogs. Because photo-$z$ kernels are large relative to the LIM channel width, the expected cross-correlation signal spans many consecutive frequency channels, which manifests as a long-wavelength mode in frequency space. Since both the filtering and cross-correlation operations are linear, the polynomial fit applied uniformly across the full frequency band will end up absorbing the target galaxy-line signal.

The galaxy overdensity maps carry no continuum emission and are never filtered, so the cleaning operator $\mathbf{M}$ of Section~\ref{sec:cib} acts on only one leg of the $g\nu$ cross-spectrum. Writing the cross-spectra at fixed $\ell$ as a vector over the $N_\nu$ LIM channels, $\mathbf{C}_\ell^{g\nu}$, the cleaned spectra are
\begin{equation}
    \mathbf{C}_\ell^{g\nu,\,\mathrm{clean}} = \mathbf{M}\,\mathbf{C}_\ell^{g\nu} \ ,
    \label{eq:gl_clean}
\end{equation}
instead of the two-sided $\mathbf{M}\mathbf{C}_\ell\mathbf{M}^\top$ of Eq.~\ref{eq:mcm}, which applies when both fields are filtered. The operator mixes channels along the frequency axis, and a mode as smooth as the photo-$z$ cross-correlation can be well described by a polynomial, so most of the signal ends up in the subtracted continuum. This over-subtraction leads to severe attenuation of the cross-spectrum and induces artificial negative correlations in the residual spectra.

\begin{figure}
    \centering
    \includegraphics[width=\linewidth]{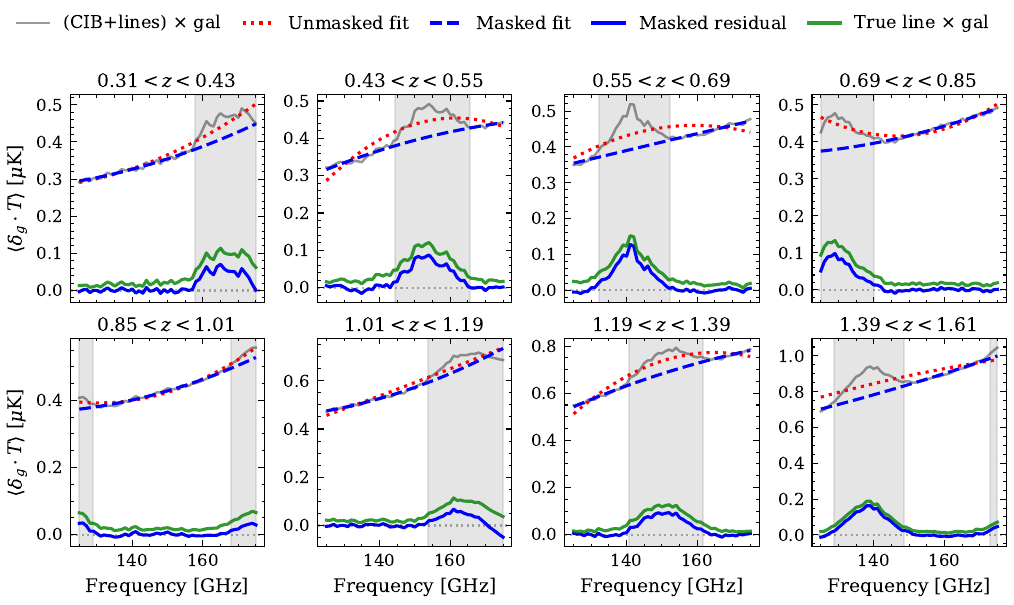}
    \caption{Continuum removal for the galaxy-line cross-spectra, shown as the correlation $\langle \delta_g T\rangle$ between a photo-$z$ galaxy map and each frequency channel of the LIM cube. Because each photo-$z$ bin is broad, its correlation with the LIM cube extends over many frequency channels, and a polynomial fit along the line of sight would partly absorb it along with the CIB. For each photo-$z$ bin, we therefore mask the channels where line emission from that redshift range falls (shaded regions) before fitting the polynomial. The masked fit (dashed blue) is anchored by the remaining channels, which carry no correlated galaxy signal, and the residual after subtracting it (solid blue) recovers the true line-galaxy correlation (green). The unmasked fit (dotted red) instead follows the line signal and removes it along with the CIB. The grey curve shows the total CIB + CO correlation before cleaning.}
    \label{fig:gl_mask}
\end{figure}

To mitigate this signal loss, we mask the expected frequency range of each target emission line prior to the polynomial fit, computed analytically for each galaxy redshift bin as shown in Figure~\ref{fig:gl_mask}. This forces the polynomial to anchor to the signal-free edges of the band and interpolate smoothly across the masked window, decoupling the broad photo-$z$ cross-correlation mode from the CIB baseline.

The effect of this masking on the recovered cross-spectra is significant, as shown in Figure~\ref{fig:gl_mask}. For the 150~GHz channel crossed with the $0.43<z_g<0.55$ bin, which picks up CO(2-1) from $z\simeq0.54$, fitting the polynomial over the full band without any masking recovers only $20-40\%$ of the true galaxy-line cross spectrum, whereas the masked fit recovers $70-90\%$. We note that the effectiveness of this masking depends strongly on the photo-$z$ binning, with the surviving amplitude ranging from $f=0.3$ to $f=0.95$ across the plotted 8 bins. Since the transfer correction inflates the $g\nu$ errors in proportion to the signal lost, both the binning scheme and the band coverage should be optimized against the final $\langle bI\rangle$ constraints.

\bibliography{bibliography}{}
\bibliographystyle{aasjournalv7}

\end{document}